\documentclass[aps,prx,amsmath,floats,floatfix,twocolumn,
  superscriptaddress,nofootinbib,showpacs]{revtex4-1}

\pdfoutput=1 
\usepackage{diagbox}
\usepackage{mathrsfs}
\usepackage{multirow}
\usepackage{braket}
\usepackage{amsfonts}
\usepackage{amsmath}
\usepackage{amssymb}
\usepackage{amsthm}
\usepackage{bm}
\usepackage{graphicx}
\usepackage{graphics}
\usepackage{latexsym}
\usepackage{rotating}
\usepackage[dvipsnames]{xcolor}
\usepackage{mathrsfs}
\usepackage{microtype}
\usepackage{verbatim}
\usepackage{url}

\usepackage[caption=false]{subfig}
\usepackage[normalem]{ulem}

\usepackage[breaklinks=true]{hyperref}
\hypersetup{
    colorlinks=true,
    linkcolor=NavyBlue,
    filecolor=Magenta,
    urlcolor=NavyBlue,
    citecolor=NavyBlue
}

\usepackage{yfonts}
\usepackage{float}
\usepackage{xspace}
\usepackage{mathrsfs}
\usepackage{siunitx}
\usepackage{array}

\allowdisplaybreaks[4]

\newcommand{\be}{\begin{equation}}
\newcommand{\ee}{\end{equation}}

\newcommand{\ba}{\begin{align}}
\newcommand{\ea}{\end{align}}

\makeatletter
\newcommand*{\rom}[1]{\expandafter\@slowromancap\romannumeral #1@}
\makeatother

\AtBeginDocument{%
    \newwrite\bibnotes
    \def\bibnotesext{Notes.bib}
    \immediate\openout\bibnotes=\jobname\bibnotesext
    \immediate\write\bibnotes{@CONTROL{REVTEX41Control}}
    \immediate\write\bibnotes{@CONTROL{%
    apsrev41Control,author="08",editor="1",pages="1",title="0",year="1"}}
    \if@filesw
    \immediate\write\@auxout{\string\citation{apsrev41Control}}%
    \fi
}

\allowdisplaybreaks

\begin{document}

\title{Ringing out General Relativity: \\ Quasinormal Mode Frequencies for Black Holes of Any Spin in Modified Gravity}

\author{Adrian Ka-Wai Chung}
\email{akwchung@illinois.edu}
\affiliation{Illinois Center for Advanced Studies of the Universe \& Department of Physics, University of Illinois Urbana-Champaign, Urbana, Illinois 61801, USA}

\author{Nicol\'as Yunes}
\affiliation{Illinois Center for Advanced Studies of the Universe \& Department of Physics, University of Illinois Urbana-Champaign, Urbana, Illinois 61801, USA}

\date{\today}

\begin{abstract} 


After black holes collide, the remnant settles to a stationary state by emitting gravitational waves. 
Once nonlinearities subside, these ringdown waves are dominated by exponentially damped sinusoids, or quasinormal modes. 
We develop a general method using perturbative spectral expansions to calculate the quasinormal-mode frequencies and damping times in a wide class of modified gravity theories for black holes with any subextremal spin. 
We apply this method to scalar-Gauss-Bonnet gravity to show its accuracy, thus enabling robust ringdown tests with gravitational wave data.

\end{abstract}

\maketitle

\noindent {\emph{The ringdown laboratory}}. 
The cataclysmic collision of black holes results in a greatly deformed black hole remnant, which settles down to its final stationary configuration through the emission of gravitational waves~\cite{Vishveshwara:1970cc}. A short time after a common apparent horizon has formed, these gravitational waves are dominated by a sum of exponentially damped sinousoids, known as quasinormal modes~\cite{Regge_Wheeler_gauge, Press:1971ApJ}. Each of these modes is characterized by an amplitude or excitation factor that encodes details about the nonlinear merger, and by a real frequency and damping time that depend only on the parameters that define the final stationary black hole. In general relativity, these parameters are the final black hole's mass and spin thanks to the no-hair theorems~\cite{NHT_01}, because any charge the progenitors could have had is expected to quickly neutralize in astrophysical environments~\cite{Blandford:1977ds}. Outside general relativity or when immersed in matter environments, the real frequencies and damping times may also depend on other parameters~\cite{QNM_dCS_01, QNM_EdGB_01, Chung_02, Cardoso:2021wlq, Cardoso:2022whc}. 
Quasinormal modes from ringdown gravitational waves are therefore an ideal laboratory to test the validity of general relativity, and to detect or constrain modified gravity effects around black holes \cite{Franchini:2023eda}.    

For ringdown gravitational waves to uphold their promise as new laboratories for fundamental physics, it is critical that we understand how black holes ring outside Einstein's theory. In general relativity, the quasinormal mode frequencies and damping times of rotating black holes can be computed using curvature perturbations in the Newman-Penrose formalism \cite{Newman:1961qr}. In the 1970s, Teukolsky used this formalism and a special symmetry of Kerr black holes to decouple the linearized Einstein equations and find a single ``master equation,'' which admits separation of variables for the perturbed Newman-Penrose scalar that characterizes gravitational radiation~\cite{Teukolsky_01_PRL}. Later, Detweiler found this master equation could be solved numerically~\cite{Detweiler:1980gk}, and Leaver found an analytical continued fraction solution~\cite{Leaver_01}, yielding the complex quasinormal frequencies of black holes with any subextremal spin.

This method to calculate the quasinormal frequencies meets severe difficulties outside general relativity or for black holes in matter environments, in part because of the loss of the (Petrov Type D) symmetry of Kerr black holes. In principle, these difficulties can be addressed with the recently developed modified Teukolsky formalism~\cite{Modified_TE_01, Modified_TE_02}; however, this framework can be difficult to implement in practice, requiring metric reconstruction and various numerical integrations of Teukolsky-like equations. Therefore, until very recently, the gravitational physics community has had to either rely on small-spin approximations~\cite{Wagle:2021tam, Pierini:2022eim} or on full numerical relativity simulations~\cite{Okounkova:2019zjf, AresteSalo:2022hua} to calculate quasinormal frequencies and damping times for black holes in modified gravity. The small-spin approximation assumes the final black hole is spinning extremely slowly relative to its maximum extremal value, so that the linearized field equations can be reexpanded in small spin and decoupled, spin order by spin order. The small-spin approximation is known to fail for black holes that spin faster than $10\%$--$20\%$ of their maximum value in modified gravity~\cite{QNM_dCS_01, Pierini:2021jxd}, rendering this scheme nonapplicable to the analysis of ringdown gravitational waves from most black hole mergers. Full numerical relativity simulations can be used to bypass this problem, by fitting the numerical ringdown stage to a set of quasinormal modes, and numerically extracting the frequencies and damping times. These calculations, however, are limited by inaccuracies related to the simulation of mergers in modified gravity, secular, and numerical errors~\cite{Okounkova:2019zjf, AresteSalo:2022hua}, and overfitting error due to contamination by merger nonlinearities~\cite{Cheung:2022rbm, Mitman:2022qdl}.  

We here present a novel method based on perturbative spectral expansions to solve for the complex quasinormal frequencies for black holes with any subextremal spin in theories beyond general relativity. 
This method is highly accurate, computationally efficient and generic, as we demonstrate by implementing it on a member of a wide class of modified gravity theories for black holes with dimensionless spin $a = |\vec{J}|/M^2 \in [0,0.85]$, where $M$ is the black hole mass and $\vec{J}$ is the spin angular momentum. We will show that the frequencies we obtain with this method are consistent with small-spin expansions~\cite{Pierini:2022eim} as long as the dimensionless spin $a \lesssim 0.3$ \cite{Chung:2024vaf}. 
For example, for $a=0.4$, the modifications to the damping time computed using our method can be different from that obtained in \cite{Pierini:2022eim} using small-spin expansions by $\geq 40 \% $.
For larger spins, small-spin expansions are inaccurate, and the frequencies computed with our method should be used instead. We determine the optimal polyomial fitting function for these frequencies, which can now be readily used in ringdown tests of general relativity with gravitational wave data. 

\vspace{0.2cm}
\noindent {\textit{A wide class of modified gravity theories}}.
Modified gravity is usually investigated to address anomalies in observations or theoretical aspects, such as the small cosmological constant that explains the observed late-time acceleration of the cosmos~\cite{late_time_acceleration_01, late_time_acceleration_02}, enhanced parity violation in the early Universe to address the observed matter-antimatter asymmetry~\cite{Gell-Mann:1991kdm}, or the incompatibility between general relativity and quantum mechanics~\cite{Calmet:2004mp}. 
To first order in deformations to general relativity, many of these theories can be described by the Lagrangian density 
\begin{equation}\label{eq:Lagrangian}
16 \pi \mathscr{L} = R - \frac{1}{2} \nabla_{\mu} \Phi \nabla^{\mu} \Phi- V(\Phi) + \alpha f(\Phi) \mathscr{Q}, 
\end{equation}
where henceforth we adopt geometric units ($c=1=G$) and the Einstein summation convention. In the above equation, $\Phi$ is a scalar field that couples to the spacetime metric, $V(\Phi)$ is a potential, $\alpha$ is a coupling constant with units of length squared that characterizes the strength of deviations from general relativity, $f(\Phi)$ is a coupling function, and $\mathscr{Q}$ is a scalar constructed from the curvature tensor. The quantities $V(\Phi)$, $\mathscr{Q}$, and $f(\Phi)$ determine the type of modified gravity theory. For example, the general sub-class of Gauss-Bonnet theories is obtained when $\mathscr{Q} = \mathscr{G} = R^2 - 4 R_{\alpha \beta} R^{\alpha \beta} + R_{\alpha \beta \gamma \delta} R^{\alpha \beta \gamma \delta}$, where $\mathscr{G}$ is the Gauss-Bonnet invariant \cite{Ripley:2019irj, Cano_Ruiperez_2019} with $R$ the Ricci scalar, $R_{\mu \nu}$ the Ricci tensor and $R_{\mu \nu \rho \sigma}$ the Riemann tensor; similarly, the subclass of dynamical Chern-Simons gravity arises when $\mathscr{Q} = \mathscr{P} = R_{\nu \mu \rho \sigma}{ }^* R^{\mu \nu \rho \sigma}$, with ${}^* R^{\mu \nu \rho \sigma}$ the dual Riemann tensor, and $\mathscr{P}$ the Pontryagin invariant~\cite{Alexander:2009tp,QNM_dCS_01}. 
Even though these two classes of theories admit a dynamical scalar field, there are only two (transverse-traceless) gravitational degrees of freedom that are propagating ~\cite{Wagle:2019mdq}.

Varying $\mathscr{L}$ with respect to the metric and the scalar field yields the field equations
\begin{align}
\label{eq:field_eqs}
& R_{\mu}{}^{\nu} + \zeta \left( \mathscr{A}_{\mu}{}^{\nu} - \bar{T}_{\mu}{}^{\nu} \right) = 0,~~~\text{and}~~~\square \vartheta + \mathscr{A}_{\vartheta} = 0, 
\end{align}
where we have set $\Phi = \alpha \; \vartheta$, defined $\zeta := \alpha^2 / {\cal{L}}^4$ as a dimensionless coupling parameter, with ${\cal{L}}$ the characteristic length scale of the physical scenario considered, and $\Box = (-g)^{-\frac{1}{2}}\partial_{\gamma} \left((-g)^{\frac{1}{2}}  g^{\gamma \lambda} \partial_{\lambda}\right)$ is the d'Alembertian operator. 
In the above equation, $\mathscr{A}_{\mu}{}^{\nu}$ is a tensor that depends on derivatives of $\vartheta$ and of $\mathscr{Q}$, $ \mathscr{A}_{\vartheta}$ is a scalar that depends on ${\mathscr{Q}}$ and derivatives of $f(\Phi)$ and $V(\Phi)$, and $2 \bar{T}_{\mu}{}^{\nu} \equiv (\nabla_{\mu} \vartheta )(\nabla^{\nu} \vartheta) + (\delta_{\mu}{}^{\nu}/\zeta) \; V(\alpha \vartheta)$ is the trace-reversed stress-energy tensor for a scalar field with a potential. 

Detected gravitational-wave signals have been used to constrain different members of this family of gravity theories, mostly by searching for the effects of modified gravity on the inspiral phase \cite{LIGOScientific:2021sio, Nair:2019iur, Perkins:2021mhb, Lyu:2022gdr}. 
For scalar-Gauss-Bonnet gravity, where $f(\Phi) = \Phi$, $V(\Phi)=0$, and $\mathscr{Q} = \mathscr{G}$, the most stringent constraint to date is $\alpha^{1/2} \lesssim 1 ~\rm km$ \cite{Nair:2019iur, Perkins:2021mhb, Lyu:2022gdr}. 
However, for other theories, such as dynamical Chern-Simons gravity, no meaningful constraints have been derived yet from gravitational waves alone, because the modifications to the inspiral phase are smaller and degenerate with spin~\cite{Nair:2019iur, Perkins:2021mhb}. 
Modified gravity theories that cannot be constrained using the inspiral alone may be probed once we develop an understanding of the ringdown.

\vspace{0.2cm}
\noindent {\textit{METRICS for Modified Gravity}}.
The basis of the ``Metric pErTuRbations wIth speCtral methodS" (METRICS) approach~\cite{Chung:2023wkd,Chung:2023zdq} is to model the ringdown through perturbations of the spacetime metric of a background black hole, where the perturbations are product decomposed in a special way and spectrally expanded. Solving the linearized field equations then reduces to a linear algebra problem for the unknown spectral coefficients and the quasinormal frequencies. This approach has been already validated in general relativity for perturbations of both Schwarzschild~\cite{Chung:2023zdq} and Kerr black holes~\cite{Chung:2023wkd}. In what follows, we extend METRICS to modified gravity.   

Current tests of general relativity suggest that, if Einstein's theory is to be modified as in Eq.~\eqref{eq:Lagrangian}, then $\zeta \ll 1$ \cite{Nair:2019iur, Perkins:2021mhb, LIGOScientific:2021sio}, 
which motivates extending METRICS as an expansion in $\zeta$. 
To leading order in $\zeta$ and using coordinates $x^{\mu} = (t, r, \chi = \cos \theta, \phi)$, the spacetime and the scalar field of a perturbed black hole in modified gravity whose background $\vartheta$ is time independent, is
\begin{equation}\label{eq:field_perts}
\begin{split}
g_{\mu \nu} & = g^{\rm GR}_{\mu \nu} + \zeta \; g^{\rm (1)}_{\mu \nu} + \epsilon \; e^{i m \phi - i \omega t} \hat{h}_{\mu \nu}, \\
\vartheta & = \vartheta^{\rm (0)} + \epsilon \; e^{i m \phi - i \omega t} \hat{h}_{\vartheta}\,,
\end{split}
\end{equation}
where $g^{\rm GR}_{\mu \nu}(r,\chi)$ is the Kerr metric, and
$g^{\rm (1)}_{\mu \nu}(r,\chi)$ and $\vartheta^{\rm (0)}(r,\chi)$ are the leading-order-in-$\zeta$ metric deformation and scalar field respectively.
The latter two can be obtained by solving Eq.~\eqref{eq:field_eqs} through controlling factors and very high-order polynomials in $r, \chi $ and $a$~\cite{Cano_Ruiperez_2019}. Therefore, the sum $g^{\rm GR}_{\mu \nu} + \zeta \; g^{\rm (1)}_{\mu \nu}$ represents a background black hole immersed in a background scalar fied $\vartheta^{(0)}$, both of which are stationary and axisymmetric. These backgrounds are perturbed by $\hat{h}_{\mu \nu}$ and $\hat{h}_{\vartheta}$, which are functions of $r$ and $\chi$, because we have factored out the time and azimuthal dependence. The quantity $\epsilon \ll 1$ is a bookkeeping parameter for the perturbations, while $m$ and $\omega$ are the magnetic mode number and the complex frequency of the quasinormal mode, respectively. 

We simplify the problem by enforcing the Regge-Wheeler gauge, which one can show is permissible in many gravity theories, including general relativity \cite{Regge_Wheeler_gauge}, Gauss-Bonnet gravity \cite{Pierini:2021jxd} and Chern-Simons gravity \cite{QNM_dCS_01}. 
In this gauge, $\hat{h}_{\mu \nu}$ is fully characterized by six functions, which we label $h_{j=1, ..., 6}$. We further define $h_7 = \hat{h}_{\vartheta}$, so that all unknown functions can be collected in $h_{j=1, ..., 7}$. 
Substituting Eq.~\eqref{eq:field_perts} into Eq.~\eqref{eq:field_eqs}, and linearizing the equations with respect to $\epsilon$, we obtain the following 11 coupled partial differential equations for $h_{j}$
\begin{equation}\label{eq:PDEs}
\sum_{\alpha = \beta = 0} \sum_{j=1}^7 \mathscr{C}_{k, \alpha, \beta, j}(r, \chi) \partial^{\alpha}_{r} \partial^{\beta}_{\chi} h_j = 0. 
\end{equation}
Here, the sum ranges over the positive integers $\alpha$ and $\beta$ up to an upper limit that depends on the gravity theory, while $\mathscr{C}_{k, \alpha, \beta, j}$ are complex functions of $(r, \chi)$ that also depend on $(M, a, m, \omega,\zeta)$, and $k \in [1, 11]$ labels the equations. 

Before we can perform a spectral expansion of $h_{j}$, we need to first take care of their divergent behavior at the event horizon and at spatial infinity due to coordinate singularities. 
We can derive a controlling factor $A_{j}$ that is asymptotic to the divergent behavior of $h_{j}$ at the horizon and at spatial infinity by requiring ingoing and outgoing wave boundary conditions and using the properties of the Killing horizon for the modified gravity background black hole \cite{Chung:2024vaf}. Using $A_{j}$, we can represent $h_j$ as 
\begin{equation}\label{eq:spectral_expansion}
h_j = A_j (r) \sum_{p=1}^{N} \sum_{\ell=1}^{N} v_{j, p, \ell} ~ \varphi_{p, \ell} (r, \chi)\,,
\end{equation} 
where $\varphi_{p, \ell} (r, \chi)$ is a complete and orthogonal spectral basis of $r$ and $\chi$, 
$N$ is the spectral order ($N\sim 20$ in this work \cite{Chung:2024vaf}), $p$ and $\ell$ label the degree of the spectral basis, and $v_{j, p, \ell}$ are constants. In this work, we use a product decomposition for the spectral basis $\varphi_{p,\ell}$, composed of Legendre polynomials in $\chi$ and Chebyshev polynomials in a compactified radial coordinate \cite{Chung:2024vaf}.

Substituting Eq.~\eqref{eq:spectral_expansion} into the left-hand side of Eq.~\eqref{eq:PDEs}, and spectrally expanding, Eq.~\eqref{eq:spectral_expansion} becomes a linear algebraic system for $v_{j, p, \ell}$. 
Using linear algebra notation (in Euclidean space), we can write this system as 
\begin{equation}\label{eq:linalg_eqn}
\mathbb{D} (\omega) \cdot \textbf{v} = \left[ \mathbb{D}^{(0)} (\omega) +\zeta \; \mathbb{D}^{(1)} (\omega) \right]\cdot \textbf{v} = \textbf{0}\,, 
\end{equation}
where $\mathbb{D} (\omega)$ is a rectangular matrix of order $11 N^2 \times 7 N^2$, whose elements are functions of $\omega$ and $(M,a,m)$, while $\textbf{v}$ is a flattened list or vector composed of all $v_{j, p, \ell}$.  
Since this equation is only valid to linear order in $\zeta$, we now search for solutions that are accurate to this same order~\cite{Chung:2023wkd, Chung:2024vaf}. 
First, we write 
\begin{equation}\label{eq:w_v_pert}
\omega = \omega^{(0)} + \zeta \; \omega^{(1)} ~~~\text{and} ~~~ \textbf{v} = \textbf{v}^{(0)} + \zeta \; \textbf{v}^{(1)},  
\end{equation}
and choose a parity to focus on, which can be selected by appropriately choosing a component of $\textbf{v}^{(0)}$ and $\textbf{v}^{(1)}$~\cite{Chung:2023wkd,Chung:2024vaf}.
Then, we insert these expressions into Eq.~\eqref{eq:w_v_pert} and solve it order by order in $\zeta$. At zeroth order, corresponding to metric perturbations in general relativity, we solve $\mathbb{D}^{(0)} (\omega^{(0)}) \cdot \textbf{v}^{(0)} = \textbf{0}$ for $\omega^{(0)}$ and all other components of $\textbf{v}^{(0)}$~\cite{Chung:2023wkd}. 
At first order in $\zeta$, we solve 
\begin{align}
\left(\tilde{\mathbb{D}}^{(1)} + \omega^{(1)} \partial_\omega \tilde{\mathbb{D}}^{(0)}\right)_{\omega^{(0)}} \cdot \textbf{v}^{(0)} + \left.\tilde{\mathbb{D}}^{(0)}\right|_{\omega^{(0)}} \cdot \textbf{v}^{(1)}  =0
\end{align}
to find
\begin{equation}\label{eq:Eigenval_pert}
\textbf{x}^{(1)} = - \mathbb{J}^{-1} \cdot \left(\left.\mathbb{D}^{(1)}\right|_{\omega^{(0)}}\cdot \textbf{v}^{(0)} \right)\,, 
\end{equation}
where $\mathbb{J}^{-1}$ is the generalized inverse of the Jacobian matrix of $\mathbb{D}^{(0)}$, namely, $\mathbb{J} =\partial_{x^{(0)}} \mathbb{D}^{(0)}$,  
and ${\textbf{x}}^{(1)}$ is a vector constructed from $\omega^{(1)}$ and the unknown components of ${\textbf{v}}^{(1)}$~\cite{Chung:2023wkd, Chung:2024vaf}. 
Equation~\eqref{eq:Eigenval_pert} is the black-hole perturbation theory equivalent of time-independent perturbation theory in quantum mechanics. 
Comparing Eq.~\eqref{eq:Eigenval_pert} to the equations for the  first-order energy shift in quantum mechanics, $\textbf{v}^{(0)}$ plays the role of the modulus square of the unperturbed eigenstate wave function, $\mathbb{D}^{(1)}$ the role of a time-independent perturbation to the Hamiltonian, and the dot product the role of the volume integral. 

\vspace{0.2cm}
\noindent {\textit{Quasinormal mode frequencies in scalar Gauss-Bonnet gravity for black holes with any subextremal spin}}. To illustrate the above modified-gravity extension of METRICS, we compute the complex frequencies for black holes with $a \in [0.005, 0.85]$ in scalar-Gauss-Bonnet gravity. This theory is a member of the modified gravity class of Eq.~\eqref{eq:Lagrangian}, defined by $f(\Phi) = \Phi$, $V(\Phi) = 0$, $\mathscr{A}_{\vartheta} = \mathscr{G}$ and
\cite{Ripley:2019irj, Cano_Ruiperez_2019}, 
\begin{align*}
\mathscr{A}_{\mu}{}^{\nu} & \equiv \delta^{\nu \sigma \alpha \beta}_{\mu \lambda \gamma \delta} R^{\gamma \delta}{}_{\alpha \beta}\nabla^{\lambda} \nabla_{\sigma} \vartheta  -\frac{1}{2} \delta_{\mu}{}^{\nu} \delta^{\eta \sigma \alpha \beta}_{\eta \lambda \gamma \delta} R^{\gamma \delta}{}_{\alpha \beta}\nabla^{\lambda} \nabla_{\sigma} \vartheta, 
\end{align*}
where $\delta^{\nu \sigma \alpha \beta}_{\mu \lambda \gamma \delta}$ is the generalized Kronecker delta. 
We implement METRICS using $N \sim 20$ spectral basis functions, and background deformations to 40th order in $a$ \cite{Cano_Ruiperez_2019}. 
We solve $\mathbb{D}^{(0)} (\omega^{(0)}) \cdot \textbf{v}^{(0)} = \textbf{0}$ via Newton-Raphson iterations for $(\omega^{(0)},\textbf{v}^{(0)})$ \cite{Chung:2023wkd}, which we then use to compute $\textbf{x}^{(1)}$ via Eq.~\eqref{eq:Eigenval_pert} with a generalized inverse to double precision. 

\begin{figure}[tp!]
\centering  
\subfloat{\includegraphics[width=\columnwidth]{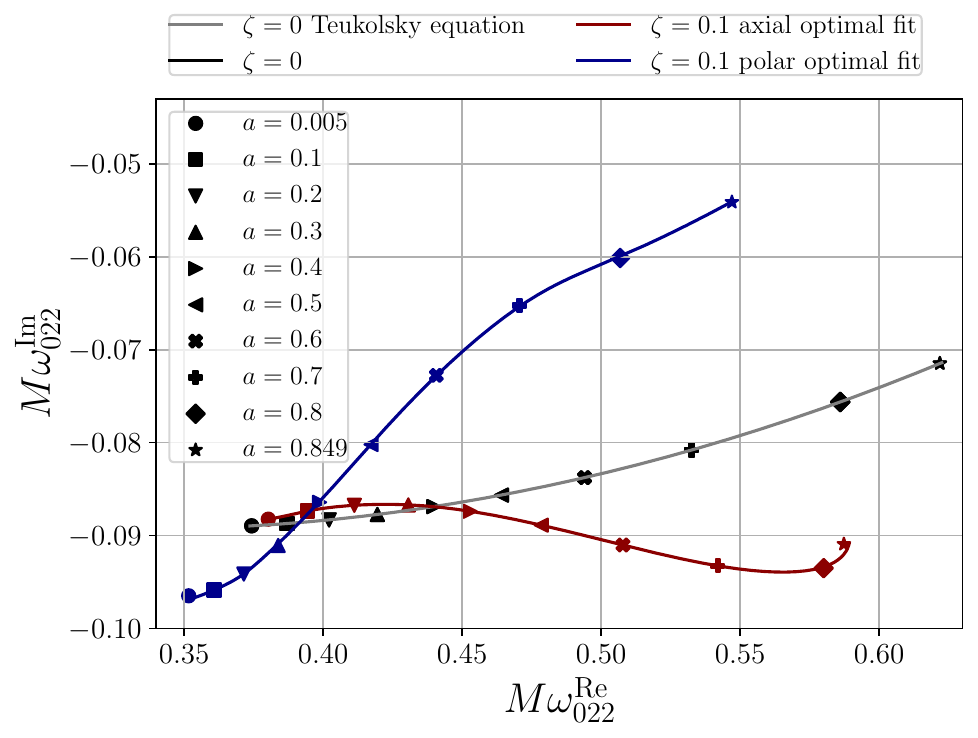}}
\caption{Quasinormal mode frequencies of axial (red) and polar (blue) metric perturbations for the 
$(n,l,m)=(0,2,2)$ mode of a black hole with $a \in [0.005, 0.85]$ in scalar-Gauss-Bonnet gravity (with $\zeta = 0.1$) and in general relativity.
Observe that the frequencies  computed by solving the Teukolsky equation with the continued fraction method in general relativity (solid gray line) agree well with the METRICS results. 
The solid red and blue lines are the trajectory of the optimal fitting polynomial of the axial and polar frequencies, respectively (see Table \ref{tab:omega_1_022}).
Observe that METRICS allows us to accurately and quantitatively explore isospectrality breaking, with the polar modes presenting the largest modifications.
}
\label{fig:omega_1_022}
\end{figure}

Figure~\ref{fig:omega_1_022} shows the complex frequencies of the axial (red) and polar (blue) metric perturbations for the fundamental corotating quadrupole mode [i.e.~$(n, l, m) = (0, 2, 2)$] for black holes with $a \in [0.005,0.85]$ and $\zeta=0$ (black) and $0.1$. 
Observe that the complex frequencies computed by solving the Teukolsky equation with a continued fraction approach \cite{Leaver:1985ax} in general relativity (solid gray line) agree with the METRICS results. 
Observe also that the axial and polar frequencies are not the same, indicating that isospectrality breaks down in scalar-Gauss-Bonnet gravity, and that the polar frequency is the most affected.
METRICS extends previous small-spin calculations~\cite{Pierini:2022eim} to large spins, and, for the first time, goes beyond purely metric modified gravity theories~\cite{Cano:2023jbk} to include axion and dilaton scalar fields. 
We have used METRICS to compute the frequencies of subdominant modes and the frequencies of the scalar field, which we will present elsewhere~\cite{Chung:2024vaf}.  

\begin{figure}[tp!]
\centering  
\subfloat{\includegraphics[width=\columnwidth]{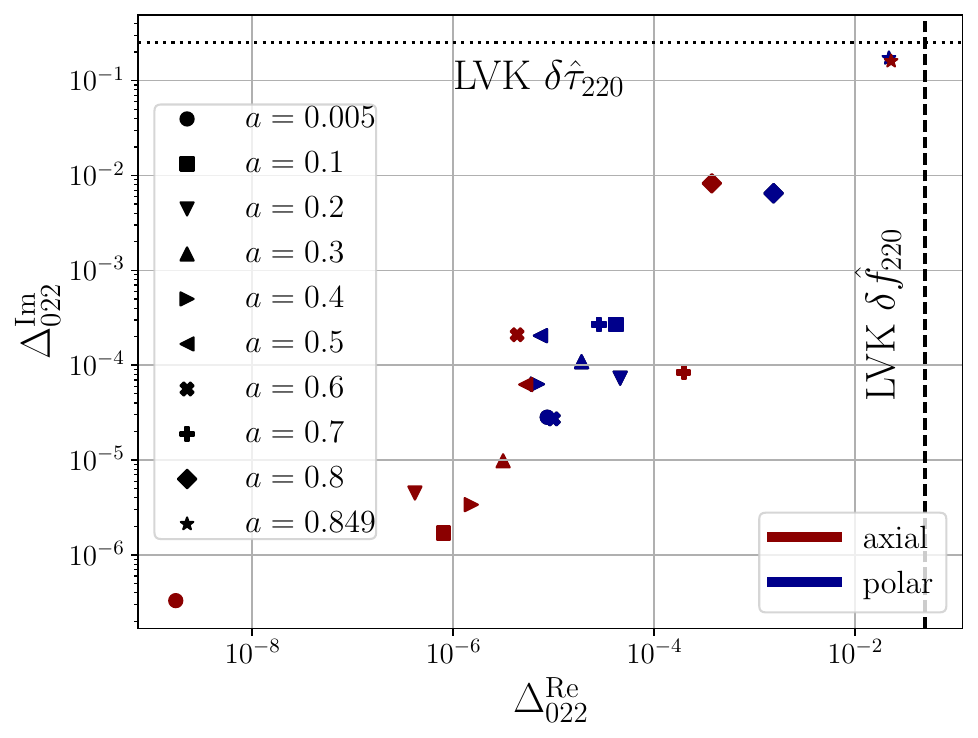}}
\caption{Numerical uncertainty of the real and imaginary parts of the frequencies of the axial (blue) and polar (red) metric perturbations, divided by the corresponding frequencies in general relativity.
The vertical (dashed) and horizontal (dotted) lines are the relative measurement uncertainty of the frequency and damping time of the fundamental corotating quadrupole mode respectively, obtained by combining all the ringdown signals detected by LIGO-Virgo-KAGRA \cite{LIGOScientific:2021sio}. 
}
\label{fig:delta}
\end{figure}

The METRICS frequencies are accurate enough for the analysis of gravitational wave ringdown data. 
Figure~\ref{fig:delta} shows the relative fractional uncertainty of the real and imaginary parts of $\omega^{(1)}$, i.e., the error in the real or imaginary part of $\omega^{(1)}$ divided the real or imaginary part of $\omega^{(0)}$, respectively\cite{Chung:2024vaf}. 
This uncertainty is smaller than $10^{-3}$ for $a \leq 0.7$, and $\sim 10^{-2}$ for $0.7 \leq a \leq 0.85$. 
Crucially, the largest uncertainty is significantly smaller than the current relative measurement accuracy of the real frequency ($\delta \hat{f}_{022} \sim 0.05$ dashed vertical line) and damping time ($\delta \hat{\tau}_{022} \sim 0.25$ dotted horizontal line) of the stacked ringdown signals detected by LIGO-Virgo-KAGRA \cite{LIGOScientific:2021sio}. 

For future parameter estimation of detected ringdown signals, we construct the optimal fitting polynomial and find that it is of degree seven and has coefficients $w_{j=0,\ldots,8}$, given in Table \ref{tab:omega_1_022}. 
The $w_0, w_1$, and $w_2$ coefficients for the polar mode are consistent with those presented in \cite{Pierini:2022eim}, despite their use of a second-order small-spin expansion, and different numerical and fitting algorithms. 
Using this fitting function, we estimate that, if a ringdown signal emitted by a remnant black hole with $a = 0.7$ in general relativity is detected by LIGO-Virgo-KAGRA with a relative measurement uncertainty of $\delta \hat{f}_{022} \sim 0.05$ \cite{LIGOScientific:2021sio}, the METRICS frequencies computed here should allow the conservative constraint $\zeta \lesssim \delta \hat{f}_{022} / \min(|\text{Re}~\omega^{(1)}_{\rm A}|, |\text{Re}~\omega^{(1)}_{\rm P}|) \sim {\cal{O}}(0.5)$, which translates to a constraint of $\sqrt{\alpha} \lesssim 12$ km for a black hole with mass $10 M_\odot$. 

\begin{table}[tp!]
\resizebox{\columnwidth}{!}{
\begin{tabular}{c|c|c}
\hline
$j$ & axial $w_j$                                   & polar  $w_j$                               \\ \hline
0   & $0.055241 + 0.00686713 i~ (9.53 \times 10^{-6} + 0.00686713 i) $ & $ -0.215202-0.0734094i ~ (0.00457066+0.00233301i) $ \\
1   & $ 0.985294+0.0636233i ~ (0.891167 + 0.0013926i)$    & $ -2.51816-0.411031i ~ (0.892279+0.271754i)$       \\
2   & $ -18.4902 + 0.325569i ~ (18.1278 + 0.029502i)$              & $ 48.4659+9.59898i ~ (19.3688+5.6713i)$          \\
3   & $ 157.802 -4.46276i ~ (146.535+0.250497i)$             & $ -431.428-82.4765i ~ (163.664+47.2939i)$            \\
4   & $ -682.224 +19.8274i ~ (609.58+1.09047i)$             & $ 1935.01+378.832i ~ (692.595+202.145i)$            \\
5   & $ 1660.08  -55.4163i ~ (1429.63+2.65319i)$         & $ -4831.15-965.058i ~ (1614.95+484.046i)$            \\
6   & $ -2307.78 + 87.744i~ (1910.19+3.64319i)$             & $ 6808.99 +1386.33i ~ (2108.76 + 656.174i)$            \\
7   & $ 1711.32 -73.0696 i~(1356.36+2.63502i)$               & $ -5065.1-1050.91i ~ (1445.54+470.048i)$         \\ 
8   & $-526.516 + 25.0324  i ~(396.962+0.779457i)$               & $ 1544.32+325.905i ~ (405.09+138.139i)$         \\ 
\hline
\end{tabular}
}
\caption{
\label{tab:omega_1_022}
Numerical values and uncertainties (in parentheses, rounded to the nearest significant decimal) of the coefficients of the optimal fitting polynomial of the axial and polar frequencies, using the fitting function $M \omega^{(1)} = \sum_{j=0}^8 w_j a^j$. 
}
\end{table}

\vspace{0.2cm}
\noindent {\textit{Future directions}}.
We have extended METRICS to enable the computation of the complex frequencies of arbitrarily spinning black holes outside general relativity. 
We exemplified this extension by applying it to scalar-Gauss-Bonnet gravity, where we showed its accuracy and efficiency, but the method is generally applicable to a wide class of theories, including Einstein-dilaton-Gauss-Bonnet gravity~\cite{Kanti:1995vq} and dynamical Chern-Simons gravity~\cite{Alexander:2009tp}.
The extended METRICS frequencies and their optimal fit are critical for ringdown tests with gravitational wave data for various reasons. 
First, the frequencies and the fits could be used directly for ringdown tests of specific theories, without incurring systematic errors due to modeling inaccuracies (e.g.~slow-rotation approximation). 
Second, the METRICS frequencies allow for mapping model-independent constraints on deviations from general relativity to constraints on the coupling constants of specific models, elevating model-independent tests beyond null-hypothesis testing. 

The extended METRICS framework can be applied to other theories of gravity and other scenarios. We have here focused on theories with massless scalar fields, where the background metric and scalar are time independent, and which only have two (transverse-traceless) gravitational wave polarizations~\cite{Wagle:2019mdq}. Nonetheless, the extended METRICS approach should be able to handle massive fields (with the appropriate modification to $V(\Phi)$ and the boundary conditions of the fields), as well as dynamical vector fields and scalar or vector polarizations, through the appropriate generalization of the field equations and of the initial choices of METRICS iterations, respectively.   
METRICS can also be applied to calculate the complex frequencies of Kerr black holes immersed in a dark matter or accretion disk environment. 
The environmental modifications to the black hole background and any dynamical environmental interactions encoded in the field equations can be treated within the extended METRICS approach. 
These are just but a few examples of the pathways that METRICS creates for future work.     

\begin{acknowledgments}
The authors acknowledge the support from the Simons Foundation through Award No. 896696, the NSF through Grant No. PHY-2207650 and NASA through Grant No.~80NSSC22K0806. 
A. K. W. C would like to thank Emanuele Berti and Dongjun Li for insightful discussion. 
N.Y. would like to thank Takahiro Tanaka for insightful discussion about this work. 
The authors would like to thank Alessandra Buonanno, Vitor Cardoso, Gregorio Carullo, Gregory Gabadadze, Leonardo Gualtieri and Shinji Tsujikawa for comments on the initial version of the manuscript.
The calculations and results reported in this Letter were produced using the computational resources of the Illinois Campus Cluster, a computing resource that is operated by the Illinois Campus Cluster Program (ICCP) in conjunction with National Center for Supercomputing Applications (NCSA), and is supported by funds from the University of Illinois at Urbana-Champaign, and used Delta at NCSA through allocation PHY240142 from the Advanced Cyberinfrastructure Coordination Ecosystem: Services $\&$ Support (ACCESS) program, which is supported by National Science Foundation Grants No. 2138259, No. 2138286, No. 2138307, No. 2137603, and No. 2138296.
The author would like to specially thank the investors of the IlliniComputes initiatives and GravityTheory computational nodes for permitting the authors to execute runs related to this work using the relevant computational resources. 
\end{acknowledgments}

\bibliography{ref}

\end{document}